\documentclass[leqno,11pt]{amsart}
\usepackage{cite} 
\newtheorem{thm}{Theorem}[section]
\newtheorem{cor}[thm]{Corollary} \newtheorem{lemma}[thm]{Lemma}
\newtheorem{prop}[thm]{Proposition}

\theoremstyle{definition}

\theoremstyle{remark}

\newtheorem{remark}[thm]{Remark} 
\newtheorem{notation}[thm]{Notation}

\newcommand{\abs}[1]{\lvert#1\rvert}

\renewcommand{\P}{\mathbb P} 
\newcommand{\R}{\mathbb R}

\newcommand{\C}{\mathbb C}
\newcommand{\Cb}{\overline{\C}} 
\newcommand{\Z}{\mathbb Z}
\newcommand{\N}{\mathbb N}

\newcommand{\ie}{i.~e.,\ }

\begin{document}
\title[zeros of orthogonal polynomials]{Zeros of  polynomials orthogonal on several intervals } 
\author{Franz Peherstorfer}
\thanks{This work was supported by the Austrian Science
Fund FWF, project-number P12985-TEC}  

\begin{abstract}
 
Let $a_1<a_2< \ldots < a_{2l}$, $E_j =[a_{2j-1},a_{2j}]$, put $E=\bigcup_{j=1}^l 
 E_j $ and 
 $H(x)=\prod^{2l}_{j=1}(x-a_j)$. Furthermore let $ p_n(x) = x^n + \ldots$ be the 
 polynomial of degree $n$ orthogonal on $E$ with respect to a weight function of
 the form $w/\sqrt{-H}$ with square root singularities at the boundary points of
 $E$ and $w\in C^3(E)$. We study and answer the following questions: how many zeros
 has $p_n$ in the interval $E_j, j\in \{1, \ldots, l\}$, when does there appear a
 zero of $p_n$ in a gap $ (a_{2j},a_{2j+1})$, are the accumulation points of the
 zeros of $(p_n)$ dense in the gaps, \ldots . First we give the number of zeros
 of $p_n$ in the interval $E_j$, $j=1,\ldots, l$, in terms of the harmonic
 measure of $E_j$ and the logarithmic mean value of the weight function. Then
 a criteria for the appearance of a zero of $p_n$ respectively of an accumulation
 point of zeros of $(p_n)$ in a gap is derived.
 As a consequence we obtain that every point of $[a_1, a_{2l}]\setminus E$ is an accumulation point
 of zeros of $(p_n)$ if the harmonic measures of the intervals are linearly
 independent over the rationals. If the harmonic measures are rational then there
 is a finite number of accumulation points of zeros of $(p_n)$ in $[a_1,
 a_{2l}]\setminus E$ only.
 \end{abstract}
   
\maketitle

\section{Introduction}
  Let $l\in {\mathbb N},$ $a_{k}\in {\mathbb R} $ for $k=1,\ldots 2l$, $a_{1}<a_{2}<\cdots <a_{2l}$,
and put  
\begin{equation*}
E_k=[a_{2k-1},a_{2k}], E=\cup _{k=1}^{l} E_k \text{ and }  H(x)=\prod 
_{k=1}^{2l}(x-a_{k}),
\end{equation*}
and set
\[
	\frac{1}{h(x)} = \begin{cases}
		(-1)^{l-k} /\pi \sqrt{-H(x)} & \text{for}\; x\in E_k, \\
		0 & \text{elsewhere}.
	\end{cases}
\]

The symbols $R$ and $S$ denote monic polynomials
that satisfy the relation
\begin{equation}
	R(x)S(x)=H(x).
	\label{eq-0}
\end{equation}
 In the following let $W$ be a function with the following properties: $W(x) \neq 0$
 on $E$, $R/(Wh)>0$ on $\operatorname{int}(E)$ and $W \in C^2(E)$ with\\
 $\lim_{n \to
 \infty} \omega_2(W,\frac{1}{n}) \ln n = 0$, where $\omega_2$ is the modulus of
 continuity of order $2$, see \cite{Lor}. By $p_n(x,R/Wh) = p_n(x) = x^n +
 \dots$, $n \in \N_0$, we denote the monic polynomial orthogonal on $E$ to
 $\mathbb P_{n-1}$ ($\mathbb P_{n-1}$ denotes as usual the set of all real
 polynomials of degree less or equal ${n-1}$) with respect to the weight function
 $R/Wh$, \ie
\begin{equation}
\label{eq-I101}
\int_E x^j p_n(x) \frac{R(x)}{W(x)h(x)}\,dx = 0 \quad\text{for } j= 0,\dots,
n-1. 
\end{equation}
$P_n(x)$ denotes the orthonormal polynomial. 

Polynomials orthogonal on several
intervals have been studied already by old masters like A. Markoff \cite{Mar},
Faber \cite{Fab}, Shohat \cite{Sho},\ldots then in the sixties and seventies
asymptotic representations of the orthonormal polynomials have been derived by
Achieser and Tomcuk \cite{Ach, ref6, tomneu}, Widom \cite{ref30} and Nuttal and
Singh \cite{Nut-Sin}, where in the last two papers also arcs and curves are
considered. Recent related results can be found in \cite{Apt, Ble-Its, Deietal2,
Ger-VanA, Ger, Mag, PehSIAM, PehJAT, Peh33, ref42, Pehfett1, Peh-Yud}.

It is well known and easy to prove by the orthogonality property \eqref{eq-I101} that
$p_n$ has all zeros in $[a_1,a_{2l}]$ and at most one zero in each gap
$[a_{2j},a_{2j+1}], j=1,\ldots,l-1$. Furthermore due to a result of Faber, see
e.g. \cite{Fab,Saf-Tot}, we have the following rough information about the
asymptotic zero distribution
\begin{equation}
	\lim_{n\to\infty}\frac{\#Z(p_n,(a_1,t))}{n} = \int_{a_1}^t d\mu_e(x)
	\label{eq-I100}
\end{equation}
where $t\in \R$, $\mu_e$ denotes the equilibrium measure of $E$, $Z(f,A)=\{x\in  
A : f(x)=0\}$ the set of zeros of $f$ on the set $A\subset \C$ and $\#Z(f,A)$ the 
number of zeros on A. Recalling (see\cite{Ran}) that 
$\mu_e(E_k)=\omega_k(\infty)$, where $\omega_k(\infty)$ denotes the harmonic 
measure for $\bar \C \setminus E$ of $E_k$ at $z=\infty$ we have moreover by \eqref{eq-I100}
\begin{equation}
	\lim_{n\to\infty}\frac{\#Z(p_n, E_k)}{n} = \omega_k(\infty) \;\text{for}\; k=1,\ldots,l.
	\label{eq-I2}
\end{equation}
In contrast to the single interval case where very detailed informations about the 
zeros are available, see e.g.\cite{Sze, Lev-Lub}, for the several interval case
such simple questions as how many zeros has $p_n$ precisely in each of the intervals
$E_k$, when does there appear a zero in a gap, is the set of zeros of $(p_n)$ dense in the
gaps are still open and will be settled in this paper. The interest in such
questions became renewed in the last years, when it turned out that orthogonal
polynomials play an important role in the solution of integrable systems, random
matrices and combinatorics, see e.g.\cite{Ble-Its, Dei1, Dei, Pas, Pas-She} and the
references therein. In many of these problems the spectrum of the associated
Jacobi operator consists of several intervals and of foremost interest are the
eigenvalues of the Jacobi operator and thus the zeros of the orthogonal
polynomials which are the eigenvalues of the truncated tridiagonal Jacobi
matrix. For the two interval case the questions on the zeros have been
investigated by A. Markoff \cite{Mar} in 1886 already. Recently we were able to
settle them completely with the help of elliptic functions \cite{Pehzellipt}.
Here we treat and settle the general case. For the special class of weights
$R/(\rho_\nu h)$, where $\rho_\nu$ is a polynomial, called Bernstein-Szeg\"o
weights and by the russian community sometimes Akhiezer weights, the precise
number of zeros in the intervals $E_k$ has been given (in somewhat weaker form)
by A. Lukashov and the author in \cite{Luk-Peh} using automorphic functions (for
the special case when $p_n$ is supposed to have all zeros in $E$ see
\cite{Kre-Lev-Nud} also).

Let us briefly outline the organization and the main results of the paper. In
the next Section polynomials orthonormal with respect to Bernstein-Szeg\"{o}
weights are studied. The main ingredient is a representation in terms of Green's
functions from which many important properties of the orthonormal polynomials
and of their zeros follow. Note,
when the weight
$R/Wh$ is approximated by Bernstein-Szeg\"{o} weights then the behaviour of the
corresponding orthonormal polynomials will be the same asymptotically as has
been shown in \cite{ref6}. In Section 3 the general case is considered and the main
results are presented. With the help of the results of Section 2 a formula for
the number of zeros in the intervals $E_k$ in terms of the harmonic measure and the mean
value of the weight function is derived as well as a criteria for the appearance of a
zero in a gap is given. Similarily as in the asymptotic description of
orthonormal polynomials one of the key stones is an associated Jacobi inversion
problem on the Riemann surface $y^2=H$. Loosley speaking, all the above desired
informations about the zeros are hidden in the solutions of the inversion
problem. For instance it turns out that a given point from a gap is an
accumulation point of zeros of $(p_n)$ if and only if the point lies in the
second (negative) sheet of the Riemann surface and is an accumulation point of
the solution of the associated Jacobi inversion problems. As a consequence we
obtain that for given $l'$, $0\leq l' \leq l-1$, points in the gaps, at most one
in each, there exists a subsequence $(n_\kappa)$ such that the points are limit
points of the zeros of $(p_{n_\kappa})$, if the harmonic measures of the
intervals $E_k$ are linearly independent over the rationals.
In the opposite case when all harmonic
measures are rational it turns out that there is a finite number of accumulation
points in the gaps only.

\section{Bernstein - Szeg\"o polynomials and their representation in terms of 
Green's functions} In the following $\rho_\nu$ denotes a real polynomial of
degree $\nu$ which has no zero in $E$, that is
\[
	\rho_\nu(x) = c \prod_{k=1}^{\nu^*}(x-w_k)^{\nu_k},
\]
where $c \in \R \setminus \{0\}$, $\nu^* \in \N_0$, $\nu_k \in \N$ for $k = 
1,\ldots,\nu^*$, $\nu = \sum_{k=1}^{\nu^*}\nu_k$, $w_k \in \C \setminus E$ for 
$k = 1,\ldots,\nu^*$, and the $w_k$ are real or appear in pairs of conjugate 
complex numbers. In this section we study polynomials, so-called 
Bernstein-Szeg\"o polynomials, orthogonal on $E$ with respect to the weight function 
$R/(\rho_\nu h)$. As already mentioned above the reason why they are important is that
polynomials orthogonal with respect to a weight function of the form $R/Wh$
behave asymptotically like those one orthogonal with respect to $R/\rho _\nu h $
if $\rho _\nu$ approximates $W$ sufficiently well.

In what follows we choose that branch of $\sqrt{H}$ for which $\sqrt{H(x)}>0$ for
$x\in(a_{2l},\infty)$. By $\phi(z,z_0)$ we denote a mapping which maps
$\bar{\C}\setminus E$ onto the exterior of the unit circle, which has a simple
pole at $z=z_0\in \Cb\setminus E$ and satisfies $|\phi(z,z_0)|\to 1$ for
$z\to\xi \in E $ a.e.; or in other words $\log |\phi| $ is the Green's function.
It is known \cite[Section 14]{ref30} that
 \begin{equation}
 	\phi(z,\infty) = \exp(\int^z_{a_1} r_\infty(\xi)\frac{d\xi}{\sqrt{H(\xi)}}),
 	\label{eq-phiinfty1}
 \end{equation}
 where $r_\infty(\xi) = \xi^{l-1} + \ldots$ is the unique polynomial such that
 \begin{equation}
 	\int^{a_{2j+1}}_{a_{2j}} r_\infty(\xi)\frac{d\xi}{\sqrt{H(\xi)}}  = 0 \quad \text 
 	{for} \, j = 1,\ldots, l-1,  
 	\label{eq-phiinfty2}
 \end{equation}
 and that for $x_0\in\R\setminus E$
 \begin{equation}
 	\phi(z,x_0) =  
 	\exp(\int^z_{a_1}\frac{r_{x_0}(\xi)}{\xi-x_0}\frac{d\xi}{\sqrt{H(\xi)}})  
 	\label{eq-phiinfty3}
 \end{equation}
 where $r_{x_0}\in\mathbb P_{l-1}$ is such that 
 \begin{equation}
 	r_{x_0}(x_0) = - \sqrt {H(x_0)}
 	\label{eq-w1}
 \end{equation}
 and
 \begin{equation}
	\mathrm{p.v.}\int^{a_{2j+1}}_{a_{2j}}\frac{r_{x_0}(\xi)}{\xi-x_0}\frac{d\xi}{\sqrt{H(\xi)}}
	= 0 \quad\text {for}\, j = 1,\ldots,l-1;
 	\label{eq-I1}
 \end{equation}
 For the following let us note that $r_{x_0}$ can be represented as 
 \[ r_{x_0}(\xi)=(\xi - x_0) \big(r_\infty(\xi) -  M_{x_0}(\xi)\big) - \sqrt{H(x_0)}, 
 \]
 where $M_{x_0}(\xi) = \xi^{l-1} + \ldots \in \mathbb P_{l-1}$ is the unique polynomial 
 which satisfies 
 \begin{equation}
	\mathrm{p.v.}\int_{a_{2j}}^{a_{2j+1}} \frac{\sqrt
	{H(x_0)}}{x_0-\xi}\frac{d\xi}{\sqrt{H(\xi)}} = \int_{a_{2j}}^{a_{2j+1}}
	M_{x_0}(\xi) \frac{d\xi}{\sqrt{H(\xi)}}\quad \text{for}\; j=1,\ldots, l-1.
	  \label{eq-int1}
 \end{equation}

\begin{lemma}
\label{lemma1}
	Let $R$, $\rho_\nu$, $\varepsilon_j \in \{-1,1\}$, $j = 1, \dots, \nu^*$, be given 
	and 
	assume that the polynomials
	$(R p_n)(x) = x^{n+\partial R} + \dots \in \mathbb P_{n+\partial R}$ and 
	$(S q_m) \in \mathbb P_{m+\partial S}$ 
	have no common zero and satisfy the relations
	\begin{equation}
		R p_n^2 - S q_m^2 = \rho_\nu g_{(n)}
 		\label{eq-210}
	\end{equation}
	with
	\begin{equation}
		{(R p_n)}^{(k)}(w_j) = \varepsilon_j ({\sqrt{H} q_m)}^{(k)}(w_j) \quad \text{for} \;\;
		k=1,\ldots,\nu _j,
		\label{eq-220}
	\end{equation}
at the zeros $w_j$, $j = 1, \dots, \nu^*$, of $\rho_\nu$ and $g_{(n)}$ a polynomial
of degree at most $l-1$ with simple zeros $x_{j,n}$ only at which, by
\eqref{eq-210},
	\begin{equation}
		R p_n(x_{j,n}) = \delta_{j,n}(\sqrt{H} q_m)(x_{j,n}) \quad \text{where}  \; \delta 
		_{j,n} \in 
		\{-1,1\}. 
		\label{eq-230}
	\end{equation}
  
	Then for $ 2 n + \partial R \ge \nu + \partial g_{(n)}$
	the following representations hold on $\C\setminus E$ 
	 \begin{equation}
	 	\mathcal R_1 := \frac{2R p_n^2}{\rho_\nu g_{(n)}} - 1  = \frac{1}{2}(\psi 
	 	_n + \frac{1}{\psi _n})
	 	\label{eq-24}
	 \end{equation}
	 
	 and 
	 \begin{equation}
		\sqrt{H}\mathcal R_2:= \sqrt{H} \frac{2q_m p_n}{\rho_\nu g_{(n)}} = \frac{1}{2}(\psi 
	 	_n- \frac{1}{\psi _n})
\label{eq-25}
\end{equation}
where
\begin{equation}
\psi_n(z) = \phi (z,\infty)^{2n+\partial R - (\nu + \partial
g_{n})}\prod^{\nu^*}_{j=1}\phi(z,w_j)^{\nu _j
\epsilon_j}\prod^{\partial{g}_{(n)}}_{j=1} \phi(z,x_{j,n})^{\delta_{j,n}}.
		\label{eq-26}
	\end{equation}
	Furthermore for $x\in E$
	\begin{equation}
		\mathcal R_1(x) = \frac{1}{2}(\psi _n^+ + \psi _n^-)
		\label{eq-260}
	\end{equation}
	and
	\begin{equation}
	    \mathcal R_2(x) = \frac{\pi}{2}\frac{(\psi _n^+ - \psi _n^-)}{ih(x)}
		\label{eq-261}
	\end{equation}
	where $\psi_n^\pm$ denotes the limiting values of $\psi _n$ from the upper and 
	lower halfplane, respectively.
	\end{lemma}

	\begin{proof} Relation \eqref{eq-210} squared can also be written in the form
		\begin{equation}
		 \mathcal R_1^2 - H\mathcal R_2^2 = 1,
			\label{eq-27}
		\end{equation}
		where $\mathcal R_1$ and $\mathcal R_2$ are given by the first expression in 
		\eqref{eq-24} and \eqref{eq-25}, respectively. Let us put on $\C\setminus E $
		\begin{equation}
			\psi_n = \mathcal R_1 + \sqrt H \mathcal R_2 = \frac{{(Rp_n + \sqrt 
			{H}q_m)}^2}{R\rho_\nu g_{(n)}}
			\label{eq-28}
		\end{equation}
		where the second equality follows with the help of \eqref {eq-210}. Again
		by \eqref{eq-27} and \eqref {eq-210}
	 \begin{equation}
			\frac{1}{\psi_n } = \mathcal R_1 - \sqrt H \mathcal R_2 =   \frac{{(Rp_n - \sqrt 
			{H}q_m)}^2}{R\rho_\nu g_{(n)}}
			\label{eq-29}
			\end{equation}
	from which it follows that the second equality holds in \eqref{eq-24} and 
	\eqref{eq-25}. 
	Since 
         \begin{equation}
		\psi_n^\pm(x) = \mathcal R_1(x) \pm  i(-1)^{l-k}\sqrt {|H(x)|}  \mathcal R_2(x) \; 
		\text{on}\; E_k 
		\label{eq-30}
	\end{equation}
	representations \eqref{eq-260} and \eqref{eq-261} follow. Note that
	\begin{equation}
		|\psi_n^\pm (x)| = 1\,\, \text{on} \,\, E.
		\label{eq-31}
	\end{equation}
	
	Further
	 \begin{equation}
	 	\begin{split}
			&\quad \psi_n \,\text{has a  pole of order}\, 2n + \partial R - (\nu +
			\partial g_{(n)})\, \text{at}\,\, z = \infty \\
	&\quad \psi_n \, \text{has a pole(zero) at}\, \, w_j \,\text{of 
	multiplicity}\, \nu_j \;\text{if}\,\,  \epsilon_j = 
	1 \;
	(\epsilon_j = -1) \\
	& \quad  \psi_n  \,\text{has a simple pole (zero) at}\, \, x_{j,n}\, 
	\text{if}\, \, \delta_{j,n} = 1 \; 
	(\delta_{j,n} = -1) \\
	\end{split}
	 	\label{eq-31E}
	 \end{equation}
	 
	Thus the function
	\begin{equation}
		f(z): = \psi_n(z) \phi(z,\infty)^{-(2n + \partial R - (\nu + \partial 
		g_{(n)})}\prod^{\nu^*}_{j=1}\phi(z,w_j)
		^{-\nu_j\epsilon_j}\prod^{\partial{g}_{(n)}}_{j=1} 
		\phi(z,x_{j,n})^{-\delta_{j,n}} 
	\end{equation}
	has neither zeros nor poles on $\bar{\C} \setminus E$ and by \eqref{eq-31} and the definition of 
 	$\phi$ we have $\abs{f^\pm} = 1$ on $E$. Hence $\log \abs{f(z)}$ is a harmonic 
 	  bounded function on $\bar{\C} \setminus E$ which has a continuous extension 
 	  to $E$ and thus $f=1$.
         Hence the 
	representation \eqref{eq-26} is proved. 
	\end{proof}
	
	\begin{remark} For the following let us note that relation \eqref{eq-210} with
	$g_{(n)}\in \P _{l-1}$ and $R/\rho _\nu h>0$ on $\operatorname{int}(E)$
	imply that $g_{(n)}$ has exactly one zero $x_{j,n}$ in each gap $[a_{2j},a_{2j+1}]$,
	$j=1,\ldots,l-1$. This follows immediately from the fact that $-RS=-H>0$, and thus by 
	\eqref{eq-210} $hg_{(n)}>0$, on $\operatorname{int}(E)$.
	\end{remark}
	
	\begin{notation} Let $t(x) = d_nx^n + d_{n-1}x^{n-1} + \ldots$ be a polynomial
	with $d_n\neq 0$. Then $\hat{t}(x) = t(x)/d_n$ denotes the monic polynomial. 
	\end{notation}

	\begin{cor}
		\label{cor1}
		Suppose that the assumptions of Lemma \ref{lemma1} are satisfied, put 
		$P_n = \sqrt{2/G_n}p_n$ and $Q_m =  \sqrt{2/G_n}q_m$, where $G_n$ denotes the 
		leading coefficient of $g_{(n)}$ and assume that $R/\rho _\nu h>0$ on $\operatorname{int}(E)$.
		Then the following statements hold for sufficiently large n:
		\begin{enumerate}
			\item[a)] On each interval $E_j$ the zeros of $RP_n$ and $SQ_m$ strictly 
			interlace.
			\item[b)] $P_n$ has a zero in $(a_{2j*},a_{2j*+1}),
			j*\in\{1,\ldots,l-1\}$, if and only if $Q_m$ has a zero in
			$(a_{2j*},a_{2j*+1})$. 
			\item[c)] $(P_{n})$ and $(Q_{m})$ are unbounded at the
			zeros of $R$ and $S$, respectively. 
%
			\end{enumerate}
			\end{cor}
			
	           \begin{proof}
				Ad a). First let us observe that by \eqref{eq-24} and \eqref{eq-25} 
			 \begin{equation}
			 	RP_n = 0\, \;\text{ iff}\;\psi_n^+ = -1\; \text{and}\; SQ_m=0\; \text{ iff}\; \psi _n^+ = 1
			 	\label{eq-34}
			 \end{equation}
			 where on $E$ $\psi_n^+$ is given in \eqref{eq-30} and naturally coincides with
			 $\psi _n$ outside $E$. By the
			 representations \eqref{eq-26} and \eqref{eq-phiinfty1} - \eqref{eq-phiinfty3} we
			 have for $x\in E $
			 \begin{equation}
			 	\psi_n^+(x) =e^{i\chi_n(x)},
				\label{eq-35}
			\end{equation}
			where 
			\begin{eqnarray*} 
				\chi_n(x) &=& -(2n+\partial R - (\nu + \partial
			g_{(n)}))\pi\int^x_{a_1} \frac{r_\infty(t)}{h(t)}\,dt \nonumber \\
				&{}&\text{+ bounded function with respect to x and n},
			 \end{eqnarray*}
		 where we have used the fact that for $x\in \operatorname{int}(E)$
			 \begin{equation}
 			 	\lim _{z\to x \text{,} Im z>0} r_\infty (z)/\sqrt{H(z)}= -i\pi r_\infty(x)/h(x). 
				\label{eq-35E}
			 \end{equation}
		 Concerning the boundedness with 
			 repect to $x$ and $n$ of the error in \eqref{eq-35} let us note first that the
			 polynomial $r_{x_0}$ from \eqref{eq-phiinfty3} is uniformly bounded with respect
			 to $x_0\in (a_{2j},a_{2j+1}), j=1,\ldots,l-1$, since the Cauchy principal-value
			 integral at the left hand side in \eqref{eq-int1} is uniformly bounded and that,
			 if $x_{j,n}$ or a subsequence tends to a boundary point of $E$,
			 \begin{equation}
			 	\lim_{n\to \infty}\int_{a_{1}}^x \frac{\sqrt{H(x_{j,n})}}{t - x_{j,n}}
			 	\frac{dt}{h(t)}
			\end{equation} 
				exists for $x\in E$, see e.g.\cite{Mus}. Since
			 	by \eqref{eq-phiinfty2} $r_\infty $ has exactly one zero in each gap
			 	$(a_{2j},a_{2j+1}), \,j=1,\ldots,l-1$, and thus $r_\infty /h >0$ on $int(E)$ it
			 	follows that for sufficiently large $n$ $\chi_n$ is strictly monotone on each
			 	interval $E_k$ which proves part a).

			 Ad b). As above by \eqref{eq-261}, \eqref{eq-26} and \eqref{eq-phiinfty1} - 
			 \eqref{eq-phiinfty3} for $x\in \operatorname{int}(E)$
			 
			 	\begin{equation}
			 	\frac{(P_{n}Q_{m})(x)}{(\rho_\nu \hat{g}_{n})(x)} = 
				\frac{\sin \Bigg(-(2n+\partial R - 
 			 	(\nu + \partial g_{(n)})\pi \int^x_{a_1} \frac{r_\infty(t)}{h(t)}\,dt + 
				\ldots \Bigg)}{(-1)^{l-k}\sqrt{-H(x)}} 
			 	\label{eq-zerosb1}
				\end{equation}
				hence, note that by \eqref{eq-34} $\sin \chi _n(a_j) = 0$ at the zeros of $H$ and that 
				$n$ is fixed,
				\begin{multline}
					\lim_{x\to a_j} \frac{(P_{n}Q_{m})(x)}{(\rho_\nu \hat{g}_{n})(x)} =\\ 
						\cos \chi _n(a_j) \Bigg(\frac{(2n+\partial R - (\nu + \partial g_{(n)})
					r_\infty(a_j) +\text{bounded function}}{H'(a_j)}\Bigg)  
					\label{eq-zerosb2}
 				\end{multline}

				Using the facts that 
				\begin{equation}
					\text{sgn}\, (r_\infty(a_{2j}) H'(a_{2j})) = \text{sgn}\,(r_\infty(a_{2j+1}
					) H'(a_{2j+1}))\;
					\text{for}\; j=1,\ldots, l-1,
					\label{eq-zerosb3}
				\end{equation}
				that $\cos \chi _n(a_j) = \pm 1$ and
				that by \eqref{eq-210} $R\rho_\nu \hat{g}_{n}>0$ on $int(E)$ 
				\eqref{eq-zerosb2} implies that
				\[ \text{sgn}\,(P_nQ_m)(a_{2j})= \text{sgn}\,(P_nQ_m)(a_{2j+1})\quad  \text{for}\; j=1,\ldots,l-1 
				\]
				if $n$ is sufficiently large. Recalling that $P_n$ and $Q_m$ have at most one zero
				in the gap part b) is proved.
				
				Ad c). The unboundedness of $(P_{n})$ and $(Q_{m})$ at the zeros of $R$ and $S$ 
				follows immediately by \eqref{eq-zerosb2}.
				\end{proof}
%
%
%

				For the following the somewhat technical statements will be needed.
				
					\begin{cor}
		\label{cor2}
		Under the assumptions and the notations of Corollary\ref{cor1} the following
		inequalities hold for $n\geq n_0$
		 \item[a)] Put $Z_\epsilon (S) = \bigcup_{a_j\in Z(S)}
		[a_j-\epsilon, a_j+\epsilon]$, $\epsilon > 0$. Then
						                           \begin{equation}
				|P_{n}|\geq const > 0 \quad \text{on} \quad Z(RQ_m,E\setminus Z_\epsilon(S))
				\label{eq-32n}
				 \end{equation}
				 For such $n$'s for which $\abs{P_n} \geq \tilde{const} > 0 $
			on $Z(S)$ we even have
			                           \begin{equation}
				|P_{n}|\geq const > 0 \quad \text{on} \quad  Z(HQ_m,E)
				\label{eq-32}
			\end{equation}
			
			\item[b)] If $|\hat{g}_{(n)}|\geq \tilde{const} > 0 $ on $E$ then the inequality
			\eqref{eq-32} as well as
			\begin{equation}
				|Q_{m}| \geq const > 0\quad \text{on} \quad Z(HP_n,E).
				\label{eq-33}
			\end{equation}
			hold.
			\end{cor}

			\begin{proof}
							Ad a). Since $\hat{g}_{(n)}$ has all zeros in the gaps it follows that
				$\abs{\hat{g}_{(n)}} \geq const > 0$ on $E\setminus (Z_{\epsilon}(S) \cup
				Z_{\epsilon}(R))$. Hence, by \eqref{eq-210}, it remains to be shown only that
				$\abs{P_n} \geq const > 0$ on $Z_\epsilon (R) \cap E_j, j \in \{1,\ldots,l\}$.
				For instance, let us consider the case that $a_{2j-1}\in Z(R)$ and $a_{2j}\in
				Z(S)$. Then by \eqref{eq-210} the following estimate holds on $E_j$
				\begin{equation}
					\abs{P_n}^2 \geq \tilde{const} \abs{\rho_\nu} \abs{\frac{x-x_{j-1,n}}{x-a_{2j-1}}}\abs{x-x_{j,n}}
					\geq  \tilde{const}\abs{\rho_\nu}\abs{x-x_{j,n}}
 					\end{equation}
					with $\tilde{const} >0$, which implies the desired inequality.
					If $\abs{P_n} \geq
					\tilde{const} > 0$ on $Z(S)$ then by \eqref{eq-210} the zeros of $\hat{g}_{(n)}$
					can not accumulate to $Z(S)$ which gives, again by \eqref{eq-210}, the second
					statement.
				
				Ad b). Follows immediately from \eqref{eq-210} and part c).
				\end{proof}

		 Next let us recall (see the author's paper \cite{PehSIAM} where a more 
		 general statement is proved even) that the polynomials which are orthonormal 
		 with respect to a measure of the form 
		 \begin{equation}
		 	d\mu _{R,\rho _\nu,\varepsilon} =\frac{R}{\rho _\nu h}dx + \sum^{\nu^*} _{j=1} 
		 	(1-\varepsilon _j) \frac{\sqrt{H(w_j))}}{\rho'_\nu (w_j)}\delta (x-w_j),
			\label{eq-pointmeasure}
		 \end{equation}
		 where $R/\rho _\nu h>0$ on $\operatorname{int}(E)$,  $\varepsilon _j \in \{-1,1\}$ for 
		 $j=1,\ldots,\nu^*$, and if $\varepsilon _j=-1$ then $w_j$ is a simple real zero of 
		 $\rho_\nu$, and where $\delta $ denotes the Dirac measure, 
		 satisfy the assumptions of Lemma~\ref{lemma1}, i.e.,\eqref{eq-210}  - \eqref{eq-230}, 
		 resp. of Corollary~\ref{cor1}. Indeed the following statement holds.

			 \begin{lemma}
				 \label{lemma2}
				 Let
				 $2n+\partial R \geq \nu + 
				 l-1$. Then $P_n \; (Q_m)$ is orthonormal to $\mathbb P_{n-1}$  
				 $(\mathbb P_{m-1})$ with 
				 respect to 
				 $d\mu _{R,\rho _\nu,\varepsilon}$  $(d\mu _{S,\rho _\nu,\varepsilon})$
				 if and only if $P_n \; (Q_m)$ satisfies a 
				 representation of the form
				 	\begin{equation}
		R P_n^2 - S Q_m^2 = 2 \rho_\nu \hat{g}_{(n)}
 		\label{eq-21neu}
	\end{equation}
	with
	\begin{equation}
		{(R P_n)}^{(k)}(w_j) = \varepsilon _j ({\sqrt{H} Q_m)}^{(k)}(w_j), 
		\quad \text{for} \;\;
		k=1,\ldots,\nu _j,
		\label{eq-22}
	\end{equation}
	at the zeros $w_j$, $j = 1, \dots, \nu^*$, of $\rho_\nu$ and $g_{(n)}$ is a polynomial of degree $l-1$ which has exactly one zero 
	$x_{j,n}$ in each gap $[a_{2j},a_{2j+1}],\, j=1,\ldots,l-1$. Moreover,
		\begin{equation}
		R P_n(x_{j,n}) = \delta_{j,n}(\sqrt{H}Q_m)(x_{j,n}) \quad \text{where}  \; \delta 
		_{j,n} \in 
		\{-1,1\}. 
		\label{eq-2302}
	\end{equation}
 \end{lemma}

		\section{Main results}
	 
		Let $\mathcal R$ denote the hyperelliptic Riemann surface of genus $l-1$ defined 
		 by $y^2 = H(z)$ with branch cuts $[a_1,a_2], 
		 [a_2,a_3],\ldots,[a_{2l-1},a_{2l}].$ The two sheets of $\mathcal R$ are 
		 denoted by $\mathcal R^+$ and $\mathcal R^-$, where on  $\mathcal R^+$ the 
		 branch of $\sqrt{H(z)}$ is chosen for which $\sqrt{H(x)}>0$\, for $x > 
		 a_{2l}$.
		 To indicate that $z$ lies on the first resp. second sheet we write $z^+$ and 
		 $z^-$. 
		 Furthermore let the cycles  $\{\alpha _j,\beta 
		 _j\}_{j=1}^{l-1} $be the usual canonical 
		 homology basis  on $\mathcal R$, i.e., the curve $\alpha _j$ lies on the 
		 upper sheet $\mathcal R^+$ of $\mathcal R$ and encircles there clockwise the interval 
		 $E_j$ and the curve $\beta _j$ originates at $a_{2j}$ 
		 arrives at $a_{2l-1}$ along the upper sheet and turns back to $a_{2j}$ along 
		 the lower sheet. $\mathcal R' $ denotes now the simple connected canonical 
		 dissected Riemann surface.  Let $\{ \varphi _1,\ldots,\varphi_{l-1}\}$, 
		 where $\varphi _j = \sum^{l-1}_{s=1} d_{j,s} \frac{z^s}{\sqrt{H(z)}}dz,
		 d_{j,s}\in\C$, be a base of the normalized differential of the first kind i.e.
		 \begin{equation}
		 	\int_{\alpha _j}\varphi _k = 2\pi i\delta _{jk} \quad \text{and}\quad
		 	\int_{\beta_j}\varphi _k = B_{jk} \quad \text{for}\; j,k = 1,\ldots,l-1
		 	\label{eq-38}
		 \end{equation}
		 where $\delta_{jk}$ denotes the Kronecker symbol here. Note that the $d_{j,n}$'s 
		 are real since $\sqrt{H(z)}$ is purely imaginary on $E_j$ and since
		 $\sqrt{H(z)}$ is real on $\R \setminus E$ the $B_{jk}$'s are also real.
		 \newline
		 In the following $\eta _3 (P,Q) $ denotes the differential of the third kind which
		 has simple poles at P and Q with residues 1 and -1, respectively, and is
		 normalized such that
		 \begin{equation}
		 	\int_{\alpha_j}\eta _3 (P,Q) dz = 0\,\quad \text{for}\quad j =1,\ldots,l-1
		 	\label{eq-39}
		 \end{equation}

		 \begin{lemma}
			 \label{lemma3.1}
			  Let $R$, $\rho_\nu$ be such that $R/\rho _\nu h>0$ on $\operatorname{int}(E)$,
			  let $\varepsilon_j \in \{-1,1\}$, $j = 1, \dots, \nu^*$, and assume that the polynomials $(R p_n)(x) = x^{n+\partial R} + \dots \in
			  \mathbb P_{n+\partial R}$ and $(S q_m) \in \mathbb P_{m+\partial S}$ have no
			  common zero and satisfy the relations
	\begin{equation}
		R p_n^2 - S q_m^2 = \rho_\nu g_{(n)}
 		\label{eq-21}
	\end{equation}
	with $g_{(n)} \in \P_{l-1}$ and 
	\begin{equation}
		{(R p_n)}^{(k)}(w_j) = \varepsilon_j ({\sqrt{H} q_m)}^{(k)}(w_j),
		\quad \text{for} \;\;
		k=1,\ldots,\nu _j,
		\label{eq-22000}
	\end{equation}
	at the zeros $w_j$, $j = 1, \dots, \nu^*$, of $\rho_\nu$. 
	Then for each $n\geq n_0$, $k=1,\ldots, l-1$,

	\begin{equation}
		\begin{split}
			\sum_{j=1}^{l-1}\delta _{j,n} \int_{x_{j,n}^-}^{x_{j,n}^+} {\varphi _k}& =
		-(2n + \partial R - (\nu + \partial g_{(n)})) \int_{\infty^-}^{\infty^+}{\varphi
		_k }\, -\sum^{\nu^*}_{j=1}\nu_j \varepsilon _j \int _{{w_j}^-}^{{w_j}^+}\varphi
		_k \\& \quad  +\sum_{j=1}^{l-1} (2\#Z(P_n,E_j) + \#Z(R,E_j))B_{kj},
	          \end{split}
 		\label{eq-40}
 		\end{equation}
		where the $x_{j,n}$'s, $x_{j,n}\in [a_{2j},a_{2j+1}]$ for $j=1,\ldots,l-1$, are
		the zeros of $g_{(n)}$ and the $\delta_{j,n}$'s, $\delta _{j,n} \in \{-1,1\}$
		for $j=1,\ldots,l-1$, are given by
			\begin{equation}
		R p_n(x_{j,n}) = \delta_{j,n}(\sqrt{H} q_m)(x_{j,n}), 
		\label{eq-23}
	\end{equation}
		that is, by relation \eqref{eq-21}, and where we integrate on $\mathcal R'$.
		\end{lemma}

	\begin{proof}
		As in Lemma~\ref{lemma1} let us put
		\begin{equation}
			\psi _n = \frac{(Rp+\sqrt{H}q)^2}{R\rho g}
			\label{eq-41}
		\end{equation}
		Note that by  \eqref{eq-21}
		\begin{equation}
			\frac{1}{\psi _n} = \frac{(Rp-\sqrt{H}q)^2}{R\rho g} 
			\label{eq-42}
		\end{equation}
	 
		Hence, in addition to \eqref{eq-31E} we have 
		 \begin{equation*}
		 	\begin{split}
			   &x= \infty^- \,\,\text{is a zero of}\,\,\psi _n \,\,  \text{of 
			   multiplicity}\, 2n + 
				\partial R - (\nu + \partial g_{(n)}) \\ 
		 &x = {w_j}^- \text{ is a pole (zero) of}\,\,  \psi _n \,\, \text{of 
		 multiplicity}\,\, \nu _j \,\,\text{if}\,\, \varepsilon _j = 
	 -1(+1)\\ 
	 &x=x_{j,n}^-\;\text{is a simple pole (zero) of}\,\,\psi_n  \,\,  
	\text{if}\, \, \delta_{j,n} = -1(+1)\, .  
	 \end{split}
	 \end{equation*}

		Now let us consider the differential
		 \begin{equation}
		 	 \eta _{(n)}:= d\ln \psi _n(z)
		 	\label{eq-43}
		 \end{equation}
		 Then it follows that $\eta _{(n)}$ has a representation of the form
		 \begin{eqnarray}
		 	\eta _{(n)} &= &(2n + \partial R - ( \nu + \partial g_{(n)})) \eta _3(\infty
		 	^-,\infty ^+) +\sum^{\nu^*}_{j=1}\nu _j \varepsilon _j\eta _3({w_j}^-,{w_j}^+)
		 	\nonumber \\ &{}& +\sum_{j=1}^{l-1}\delta _{j,n} \eta _3(x_{j,n}^-,x_{j,n}^+)
		 	+ \sum_{j=1}^{l-1} c_{j,n}\varphi _j
			\label{eq-44}
		 \end{eqnarray}
		 where $c_{j,n}\in\C$. By the normalization  \eqref{eq-39} of $\eta _3$ we obtain
	 
	\begin{equation}
		 	\int _{\alpha _j} \eta _{(n)} = \sum_{k=1}^{l-1}c_{k,n} \int_{\alpha _j}\varphi 
		 	_k = 2\pi ic_{j,n}
		 	\label{eq-45}
		 \end{equation}

		 On the other hand it follows by 
		 shrinking $ \alpha _j$ to $E_j$ and by $|\psi _n(z)| = 1 $ on $E_j$ that 
		 \begin{equation}
			\int _{\alpha _j} \eta _{(n)} = -i\Delta_{E_j}arg\psi _n = -2\pi i(2\#Z(P_n,E_j)
			+ \#Z(R,E_j))
		 	\label{eq-46}
		 \end{equation}
		 where the last equality follows with the help of  \eqref{eq-34} and the facts 
		 that by 
		 \eqref{eq-30} $arg \psi _n^- = -arg \psi _n^+$ and that we have shown in the 
		 proof of Corollary \ref{cor1} a) that $\chi _n = arg\psi_n^+$ is strictly 
		 monotone for sufficiently large $n$. Since by the bilinear relation for abelian differentials of the first 
		 and 
		 third kind (see in particular \cite[pp.394-402]{Osg} or \cite{Spr})
		 \begin{equation}
		 	\int _{\beta _k}\eta _3(P,Q) = \int_P^Q\varphi _k
		 	\label{eq-47}
		 \end{equation}
		 we get, recall \eqref{eq-38}, 
		 \begin{eqnarray}
			 	 \int_{\beta _k} \eta _{(n)} &=& (2n + \partial R - (\nu + \partial g_{(n)}))
			 	 \int_{\infty^-}^{\infty^+}{\varphi _k }\, +\sum^{\nu^*}_{j=1}\nu _j \varepsilon
			 	 _j \int _{{w_j}^-}^{{w_j}^+}\varphi _k \nonumber\\
	          &{}&+ \sum_{j=1}^{l-1}\delta _{j,n} \int_{x_{j,n}^-}^{x_{j,n}^+} {\varphi _k} - 
		 \sum_{j=1}^{l-1} (2\#Z(P_n,E_j) + \#Z(R,E_j))B_{kj} 
		 	\label{eq-48}
		 \end{eqnarray}
		 Now $\psi_n$ takes the same value at both sides of the cross point of $\beta _k$
		 and $\alpha _k$ and thus
		 \begin{equation}
		 	\int_{\beta _k} \eta _{(n)} = 2m_k\pi i ,
		 	\label{eq-49} 
		 \end{equation}
		 where $m_k\in\Z$. Since, by the Remark before \ref{cor1} the $x_{j,n}$'s lie in
		 the gaps and the $w_j$'s are real or appear in pairs of complex numbers, the right
		 hand side of \eqref{eq-48} takes on real values, recall that we integrate on the
		 simple connected Riemann surface $\mathcal R$', and thus $m_k=0$ which is the
		 assertion.
	 \end{proof}

		 Let us note that another version of Lemma \ref{lemma3.1} has been proved by A.
		 Lukashov and the author in \cite{Luk-Peh} with the help of automorphic
		 functions.
		  \newline
		  Now we are able to improve Faber`s result \eqref{eq-I2}. In fact under some mild
		  additional conditions we obtain a very presice estimate of the number of zeros
		  in the intervals in terms of the harmonic measure and the mean value of the
		  weight function. For the following let us recall that
		  every $\vec{v}\in\C^{l-1}$ can be
		  represented in the form $\vec{v} = 2\pi i\vec{\mu} + B\vec{\lambda} + 2\pi i
		  \vec{n} + B\vec{m}$, where $\vec{n},\vec{m} \in \Z^{l-1}$ and $\mu _j,\lambda _j
		  \in [0,1)$ for $j=1,\ldots,l-1$. The quotient space $\C^{l-1}/(2\pi i\vec{n} +
		  B\vec{m})$, $\vec{n},\vec{m} \in \Z^{l-1}$, called the Jacobi variety of the
		  surface $\mathcal R$ and denoted by $ \mathcal Jac \mathcal R $, is a $2(l-1)$
		  -dimensional real torus. 
		  
		  					  		 \begin{thm}
			 \label{Nst.Thm.1}
			 Let $W \in C^2(E)$ with $\lim_{n \to \infty} \omega_2(\frac{1}{n}) \ln n = 0$ 
                           $W \neq 0$ on $E$ and 
						   $R/Wh>0$ on $int(E)$. Then the following statements hold:
						   \begin{enumerate}
							   \item[a)] For all $n\geq n_0$, $j=1,\ldots,l$,
							   \begin{equation}
							   	\# Z(P_n(.,R/Wh),E_j) = n\omega _j(\infty) + O(1),
							   	\label{eq-Nhm1}
							   \end{equation}
							   where $\omega_j(\infty)$
						   denotes the harmonic measure of the interval $E_j$.
						   \vspace{0.1cm}
							   \item[b)] 
							   Put $\vec{\varphi}(W) =
                           (\frac{1}{\pi i}\int _E \varphi _j^+ \log |W|)_{j=1}^{l-1}$  and let \\
                           $\vec{\mathcal V}_n = (2n+\partial R - l+1)\frac{\vec{\omega}(\infty)}{2} -
                           B^{-1}\vec{\varphi}(W) - \frac{\# \vec{Z}(R)}{2}$. Then
			for the $n$'s, $n\geq n_1$, for which $\abs{P_n(.,R/Wh)} \geq \tilde{const} > 0$
		at the zeros of $S$, the inequalities
			\begin{equation}
				\abs{\# Z(P_n(.,R/Wh),E_j) - [\mathcal{V}_{j,n} + \frac{1}{2}]}\leq 1,
				\label{eq-Nhm2}
			\end{equation}
			$j=1,\ldots,l-1$, hold. As usual, $[ . ]$ denotes the greatest integer not larger
			than . .
		\end{enumerate}
			\end{thm}
			
						\begin{proof}. First let us prove that for weights of the form $W=\rho_\nu$
			the difference between $\# Z(P_n(.,R/Wh),E_j)$ and $[\mathcal{V}_{j,n} +
			\frac{1}{2}]$ is at most one for all $n\geq n_0$, $j=1,\ldots,l-1$. It's known,
			see e.g. \cite{Nut-Sin} and for a detailed proof \cite{Pehman}, that
		\begin{equation}
	           -\frac{2}{\pi i}\int_E \varphi _k^+  \log \abs{\hat{\rho} _\nu} = 
		\sum^{\nu^*}_{j=1}\nu_j \Bigg(\int _{{w_j}^-}^{{w_j}^+}\varphi _k - \int_{\infty^-}^{\infty^+}\varphi
		_k\Bigg)
		\label{eq-neu}
		\end{equation}
		and thus relation \eqref{eq-40} becomes

	 \begin{multline}
		\label{eq-Nst7E}
		\sum_{j=1}^{l-1}\delta _{j,n}\int_{x_{j,n}^-}^{x_{j,n}^+} {\varphi _k} =
		 \frac{2}{\pi i}\int_E \varphi _k^+  \log{|\hat{\rho} _\nu|}
		 	- (2n + \partial R -  \partial g_{(n)})\int_{\infty^-}^{\infty^+}{\varphi 
		_k } \\
                  +\sum_{j=1}^{l-1} (2\#Z(P_n,E_j) + \#Z(R,E_j))B_{kj} 
               \quad \text{for}\quad  k=1,\ldots,l-1
	\end{multline}
	
	Let us put $z_{j,n}:= x_{j,n}^{\delta_{j,n}}$, where $x_{j,n}^{\pm 1}:=
	x_{j,n}^{\pm}$. Hence the points $z_{j,n}$ and
	$z_{j,n}^*=x_{j,n}^{-\delta_{j,n}}$ lie above each other on $\mathcal R$. Now the left
	hand side of \eqref{eq-Nst7E} can be represented in the form
 	 \begin{equation}
 		 \sum_{j=1}^{l-1}\delta _{j,n}\int_{x_{j,n}^-}^{x_{j,n}^+} {\varphi _k}=
		 \sum_{j=1}^{l-1}\int_{z_{j,n}^*}^{z_{j,n}} \varphi _k=\sum_{j=1}^{l-1} (\lambda_{j,n}+m_{j,n})B_{jk}
 		 \label{eq-Nst7EE}
 		 	\end{equation}
 			where $\lambda_{j,n} \in (-1,1)$ and $m_{j,n} \in \Z$, $j=1,\ldots,l-1$. In view
 			of \eqref{eq-44}-\eqref{eq-48} we have that $m_{j,n}=0$ for
 			$j=1,\ldots,l-1$. Dividing \eqref{eq-Nst7E} by $n$ and taking the limit as
 			$n\to\infty$ gives
	\begin{equation}
		\int_{\infty^-}^{\infty^+}{\varphi 
		_k } = \sum_{j=1}^{l-1} B_{kj}\omega_j(\infty)
		\label{eq-Nst8}
	\end{equation}
	where we have used the known fact (see e.g.\cite{Fab,Ran}) that
	\begin{equation}
		\lim_{n\to\infty}\frac{Z(P_n,E_j)}{n} = \omega _j(\infty) 
		\label{eq-Nst9}
	\end{equation}
	Thus \eqref{eq-Nst7E} 
	 can be written as 
	 
	 \begin{equation}
		\label{eq-Nst11}	 
	\frac{\vec{\lambda}_n}{2} - B^{-1}\vec{\varphi}(\rho_\nu) +
	(2n+\partial R - \partial g_{(n)})\frac{\vec{\omega}(\infty)}{2} - \frac{\#\vec{Z}(R)}{2} = 
	\#\vec{Z}(P_n). 	
	\end{equation}
	which proves our claim. 
	
	To prove part b) of the theorem for weight functions of the form $R/Wh$, let us first
	recall, see \cite{ref6}, that there is sequence of $\rho_\nu$'s such that
	\begin{equation}
		\vec{\varphi}(\rho_\nu) = \vec{\varphi}(W)
		\label{eq-Nst12}
	\end{equation}
	and
	\begin{equation}
		|\frac{W(x)}{\rho_\nu(x)} - 1| \leq  \frac{const.}{\nu^2}\omega _2(\frac{1}{\nu})
		\label{eq-Nst12Sch}
	\end{equation}
	where  $\omega _2$ denotes the modulus of continuity of second order, 
	and that for given $\epsilon$ there exists a $\nu_0$ such that for $\nu\geq \nu_0$  
	and for $n\geq 2 \nu$
	\begin{equation}
		P_n(z,R/Wh) = P_n(z,R/\rho_\nu h) + O(\epsilon _n) \quad \text{on}\, E \; 
		\text{with}\; |\epsilon_n| \leq \epsilon . 
	               \label{eq-Nst13} 
	\end{equation}
	Hence, by the assumption, \eqref{eq-Nst13} and Corollary\ref{cor2}a) for $\nu
	\geq \nu_0$, $n\geq 2\nu$,

  \begin{equation}
 	|P_n(x,R/\rho_\nu h)| \geq const > 0 \quad \text{on} \quad   Z(HQ_m(x,R/\rho_\nu h),E).
 	\label{eq-Nst15}
 \end{equation}
 Now by \eqref{eq-Nst13} again the sign of the two orthogonal polynomials
 $P_n(x,R/\rho_\nu h)$ and $P_n(x,R/Wh)$ at the boundary points of $E$ is the
 same, if $n$ is sufficiently large. Since each of the both polynomials has at
 most one zero in each gap they have the same number of zeros in each gap.
 Furthermore by Corollary~\ref{cor1}a) $P_n(x,R/\rho_\nu h)$, and thus by
 \eqref{eq-Nst13} $P_n(x,R/Wh)$ also, has different sign on two consecutive zeros
 of $SQ_m(x,R/\rho _\nu h)$ lying in an interval $E_j$ which implies
                       \begin{equation}                                                  
                      \#Z(P_n(.,R/\rho_\nu h), E_j) = \#Z(P_n(.,R/Wh),E_j)\, \quad \text{for} \, j =
                      1,\ldots, l-1
                       \label{eq-Nst14}                                                 
                      \end{equation}  
		which gives by \eqref{eq-Nst12} the assertion.
		
		Finally let us prove part a). The problem is to estimate the number of zeros of
		$P_n(.,R/Wh)$ in the neighbourhood of a boundary point of $E_j$ which is a zero
		of $S$ because $P_n(.,R/Wh)$ may accumulate to zero there and thus the
		asymptotics \eqref{eq-Nst13} gives no information now. The problem can be
		settled by switching to the polynomials $P_m(.,S/Wh)$. Recall that
		$S$ is given by \eqref{eq-0}.
		
		It is known, see e.g. \cite{Uva}, that $P_n(.,R/Wh)$ can be expressed in terms of
		$P_j(.,S/Wh)$, essentially by Christoffel's transformation, as follows: Put $\kappa =
		\partial R+\partial S$ then
		\begin{equation}
			\label{eq-1311}
			\begin{split}
				S(x)P_n(x,R/&Wh)  =\sum _{j=0}^\kappa \mu _{j,n} P_{n+\partial S-j}(x,S/Wh) \\
				                 &=A_{\kappa -1,n}(x) P_{n-\partial R +1}(x,S/Wh) - B_{\kappa -
				                 2,n}(x)P_{n-\partial R}(x,S/Wh)
								  \end{split}
								   \end{equation}
								  where $\mu _{j,n} \in \R$ and $A_{\kappa -1,n}$ and $B_{\kappa -
				                  2,n}$ are polynomials of degree less or equal $\kappa -1$. The second equality follows by using the recurrence relation of
				                  the $P_n$'s succesively. Thus by \eqref{eq-1311}, using the well known
				                  interlacing property of the zeros of two consecutive orthogonal polynomials,
				                  there is at least one zero of $SP_n(.,R/Wh)$ or $A_{\kappa -1,n}$ between two
				                  consecutive zeros of $P_{n-\partial R}(.,S/Wh)$. Hence, loosley speaking
				                  $P_n(.,R/Wh)$ has at most $2\kappa$ "free zeros'' on $[a_1,a_{2l}]$, \ie, zeros
				                  which may lie in any $E_j$ or in a gap. Now let us consider an interval $E_j =
				                  [a_{2j-1},a_{2j}]$ with $a_{2j}$ is a zero of $S$ and $a_{2j-1}$ a zero of $R$.
				                  We have
						\begin{equation}
							\begin{split} &\#Z(P_n(.,R/Wh),E_j) \\ &\geq \#Z(P_n(.,R/Wh),[a_{2j-1},c]) +
							\#Z(P_{n-\partial R}(.,S/Wh), [c,a_{2j}]) - 2\kappa \\ &\geq \#Z(P_n(.,R/\rho_\nu
							h),[a_{2j-1},c]) + \#Z(P_{n-\partial R}(.,S/\rho_\nu h), [c,a_{2j}]) - (2\kappa +
							4)\\&\geq \#Z(P_n(.,R/\rho_\nu h),E_j) - (4\kappa +4)
							\end{split}
						\label{eq-neu2}
						\end{equation}
						where in the second inequality we have used Corollary \ref{cor1}a) and \eqref{eq-32n},
						applied to $R$ and $S$, in conjunction with \eqref{eq-Nst13} and in the third one
						\eqref{eq-1311} again. Summing up the zeros of all intervals $E_j$ it follows by a
						very rough estimate that $P_n(.,R/Wh)$ has on $[a_1,a_{2l}]$ at most $l(4\kappa
						+ 4) + l-1$ ``free zeros''. Thus the number of zeros of
						$P_n(.,R/\rho_\nu h)$ and $P_n(.,R/Wh)$ on $E_j$ differ by a constant at most
						which gives the assertion by the first proved claim.
						\end{proof}

						In connection with the second statement of the above Theorem let us note that
						the closure of the set $\{(|P_n(s_1, R/Wh)|,\ldots, |P_n(s_{\partial S}, R/Wh)|):
						n\in \N\}$, where $s_1,\ldots,s_{\partial S}$ denote the zeros of $S$, is of the
						form $\sf{X}_{j=1}^{\partial S} [0,\eta_j], 0<\eta _j\leq \infty$, if the
						harmonic measures $\omega _j(\infty)$, $j=1,\ldots,l-1$, are linearly
						independent over the rationals. Indeed with the help of Kronecker's Lemma it
						follows - see the proof of Theorem \ref{DThm} - that the zeros $x_{j,n}$ of
						$g_{(n)}$ are dense in the gaps which gives in conjunction with \eqref{eq-21neu}
						and \eqref{eq-Nst13} the assertion.
						
						For the important class of weight functions which vanish at the boundary points
						of $E$ Theorem \ref{Nst.Thm.1}b) becomes

						\begin{cor}
							\label{cor3}
							For all $n \geq n_0$ and $j=1,\ldots,l-1$ the difference between the number of
							zeros of $P_n(.,H/Wh)$ on $E_j$ and $[\mathcal{V}_{j,n} + \frac{1}{2}]$ is at 
							most one. 
							\end{cor}
							
							To get informations about the appearance of the zeros in the gaps and other
							informations on the zeros we will have to study the behaviour of the solutions
							of the so-called real Jacobi-inversion problem. The map \newline \hspace{0.3cm}
							$\mathcal A : \mathcal R^{l-1}_{symm}\,\to\, \mathcal Jac \mathcal R $,
							$(z_1,\ldots,z_{l-1}) \to (\sum_{j=1}^{l-1} \int_{e_j}^{z_j} \varphi_1,
							\ldots,\sum_{j=1}^{l-1} \int_{e_j}^{z_j} \varphi_{l-1})$, \newline is the
							so-called Abel-map. Here $\mathcal R^{l-1}_{symm}$ denotes the $(l-1)$th
							symmetric power of $\mathcal R$ and $e_1, \ldots, e_{l-1}$ are given points on
							$\mathcal R$ . It's known that $\mathcal A$ is locally biholomorphic but not
							injective globally. The problem for given $\vec{v} = (v_1, \ldots, v_{l-1}) \in
							\mathcal Jac \mathcal R$ to find $(z_1,\ldots,z_{l-1}) \in \mathcal
							R^{l-1}_{symm}$ such that
		 \begin{equation}
			 \sum_{j=1}^{l-1} \int_{e_j}^{z_j} \varphi_k  
			 = v_k\quad \mod \text{periods}
			\label{eq-50}
		 \end{equation}
		 is called the Jacobi-inversion problem. We will be interested in the real Jacobian
		 inversion problem, that is, when the $v_k$'s are real. We need some
		 notation.

		   \begin{notation} Let $[a_{2j},a_{2j+1}]^+ , [a_{2j},a_{2j+1}]^- $ denote the two copies of \\
			   $[a_{2j},a_{2j+1}]$, 
			 $j=1,\ldots,l-1$ in $\mathcal R^+$ and $\mathcal R^-$, respectively, and let\\ 
			 $\mathcal I_j = [a_{2j},a_{2j+1}]^+ \cup [a_{2j},a_{2j+1}]^- \subset 
			 \mathcal R $ which is a closed loop on $\mathcal R$.
			 Furthermore put $\Xi =  \{ (\delta _1,\ldots,\delta _{l-1}) : \delta _j \in 
			 \{ -1, +1\}\}$ and   
			 for  $\delta \in  \{ -1, +1\}$ set
			  \begin{equation*}
				  \begin{aligned}
					  (a_{2j},a_{2j+1})^\delta =\begin{cases}
					   &(a_{2j},a_{2j+1})^+\subset \mathcal R^+ \quad  \text{if}\: \delta = 1, \\
				 &(a_{2j},a_{2j+1})^-\subset \mathcal R^-  \quad \text{if}\:\delta = -1.
					  \end{cases}
					  \end{aligned}
					  \end{equation*}
					  By $z$ and $z^*$ we denote the points which lie above each other on $\mathcal
					  R$, that is, have the same image under the canonical projection $\pi$ of
					  $\mathcal R$ on the Riemann sphere. 
					  Finally denote by $\mathcal Jac \mathcal R/_\R := \R^{l-1}/B\vec{m}$ the
					  Jacobi variety restricted to the reals.
					  \end{notation}
					  As mentioned in \cite{Deietal, Mag, Nut-Sin} the real Jacobi inversion problem
					  is uniquely solvable. More precisely
	
		 \begin{lemma}
			 \label{lemma3.2}
			 \begin{enumerate}
				 \item[a)]
			 The restricted Abel map
			\begin{equation}
				\begin{split}
		            \mathcal A : &\sf{X}^{l-1}_{j=1}\mathcal I _j\quad \to 
				 \quad \mathcal Jac \mathcal R/_\R \\
				             &(z_1,\ldots,z_{l-1})\;\to \; \frac{1}{2}(\sum^{l-1}_{j=1} \int^{z_j}_{z_j*}\varphi
				             _1,\ldots,\sum^{l-1}_{j=1} \int^{z_j}_{z_j*}\varphi _{l-1})
			\label{Abel-map}
			\end{split}
			 \end{equation}
                            is a bijection.  
			 \item[b)]
 			 For each of the $2^{l-1} \vec{\delta}$'s from $\Xi$ put $ C_{\vec{\delta}} :=
 			 \mathcal A \left( \sf{X}^{l-1}_{j=1} (a_{2j},a_{2j+1})^{\delta _j}\right)$. Then
 			 \\
			 $\cup_{\vec{\delta} \in \Xi}\bar{C}_{\vec{\delta}} = \mathcal Jac \mathcal
			 R/_\mathbb R$ ,where $\bar{C}_{\vec{\delta}}$ denotes the closure of
			 $C_{\vec{\delta}}$.
			  \end{enumerate}
			 \end{lemma}
			 
			 \begin{proof}
				 Ad a). The simplest way to prove part a) is to change the model of the Riemann
				 surface by choosing now the canonical homology basis $\{\alpha'_j,
				 \beta'_j\}_{j=1}^{l-1}$, where the curve $\alpha'_j$ originates at $a_{2j}$
				 arrives at $a_{2j+1}$ along the upper sheet and turns back to $a_{2j}$ along the
				 lower sheet and $\beta'_j$ lies in the upper sheet and encircles clockwise the
				 intervall $[a_1,a_{2j}]$, and then to proceed as in \cite{Kre-Lev-Nud}. Note
				 that the $\alpha'$ and $\beta'$ periods can be expressed easily in terms of
				 the $\alpha$ and $\beta$ periods from \eqref{eq-38}. Another possibility, see
				 \cite{Deietal}, is to prove by a van der Monde argument that the derivative is
				 nonsingular on $\sf{X}^{l-1}_{j=1}\mathcal I_j$ (which is homeomorphic to a
				 $l-1$ dimensional torus) and thus $\mathcal A$ is open. By continuity it follows
				 that $\mathcal A$ is closed and thus surjective. Since $z_j \in \mathcal I_j$
				 for $j=1,\ldots,l-1$ among $z_1z_2\ldots z_{l-1}$ there are no points which lie
				 above each other or in other words $z_1z_2\ldots z_{l-1}$ is a so-called
				 nonspecial divisor which implies by Lemma 4.3 in \cite{Nut-Sin} or IX.7 and
				 X.3 in \cite{Kra} that $\mathcal A$ is injective also. The third possibility is
				 (see \cite{Mag}) to look at the solution of the Jacobi inversion problem in
				 terms of Thetafunctions \cite[p.142]{Lan}. Indeed, by the representation it
				 follows that the solutions $z_j$, $j=1,\ldots,l-1$, have to lie in $\mathcal
				 I_j$, $j=1,\ldots,l-1$ if all $v_k$ are real from which the assertion can be
				 derived easily now.

				 Since  $\mathcal A$ is holomorphic also part b) follows immediately. 
				 \end{proof}

		 \begin{thm}
			 \label{Nst.Thm.}
			 Suppose that $W$ satisfies the same assumptions as in Theorem \ref{Nst.Thm.1} and let
			 $\vec{\varphi}(W)$ and $\vec{\mathcal V}_n$ be defined as there.
			 Then the following statements hold:
							  
				 For all $n\geq n_0$ for which $2B\vec{\mathcal V}_n\in 
				  \{B(\vec{\lambda}+\vec{m}):\vec{\lambda}\in [\epsilon, 1-\epsilon]^{l-1},
				  \vec{m}\in \Z^{l-1}\}, \epsilon >0$, the following relations hold:
				  
				  \begin{enumerate}
					  \item[a)]
				\begin{equation}
				\#Z(P_n(.,R/Wh),E_j) = [ \mathcal{V}_{j,n} + \frac{1}{2}]\quad \text{for} \quad
				j=1,\ldots,l-1.
			 	\label{eq-Nst0}
				 \end{equation}
				 \item[b)]
			 If $\vec{\mathcal V}_n \in C_{\vec{\delta}}, \vec{\delta} \in \Xi $, then 
			 \begin{equation}
			 	\# Z\big(P_n(.,R/Wh),(a_{2j},a_{2j+1}) \big) = \frac{(1-\delta_{j,n})}{2}\quad
			 	\text{for} \quad j = 1,\ldots, l-1,
				\label{eq-Nst-1}
			 \end{equation}
%
%
%
%
			 \end{enumerate}
			  \end{thm}
			  \begin{proof}
				  Let $W=\rho_\nu$. In the proof of Theorem~\ref{Nst.Thm.1} we have shown that the
				  zeros $x_{j,n}$ resp. the $z_{j,n}$ defined after \eqref{eq-Nst7E} are solutions
				  of the Jacobi-inversion problem \eqref{eq-Nst7E}, taking into consideration
				  \eqref{eq-Nst7EE}. Since by assumption $2B\vec{\mathcal{V}}_n \in
				  \{B(\vec{\lambda}+\vec{m}):\vec{\lambda}\in [\epsilon, 1-\epsilon]^{l-1},
				  \vec{m}\in \Z^{l-1}\}, \epsilon >0$, it follows first of all by \eqref{eq-Nst11}
				  that
				  \begin{equation}
					 \#Z(P_n(.,R/h\rho _\nu),E_j) = [\mathcal{V}_{j,n} + \frac{1}{2}] \quad
					 \text{for} \quad j=1,\ldots,l-1
				  	\label{eq-neu3}
				  \end{equation}
				  and furthermore  with the help of Lemma \ref{lemma3.2} and continuity
				  arguments that the $z_{j,n}$'s from \eqref{eq-Nst7EE} satisfy for $n\geq n_0$
				  $z_{j,n} \in [a_{2j}+\tilde{\epsilon},a_{2j+1}-\tilde{\epsilon}]^\pm$,
				  $\tilde{\epsilon}>0$, for $j=1,\ldots,l-1$, that is, the projection
				  $\pi(z_{j,n})=x_{j,n} \in [a_{2j}+\tilde{\epsilon},a_{2j+1}-\tilde{\epsilon}]$.
				  Hence $\abs{\hat{g}_{(n)}} \geq \tilde{const} >0$ on $E$. Now part a) follows
				  with the help of Corollary\ref{cor2}b) and relation \eqref{eq-Nst13} by the same
				  arguments used in the second part of the proof of Theorem\ref{Nst.Thm.1}. Let us 
				  note for the next step of the proof that in particular the absolute values of
				  $P_n(.,R/\rho_\nu h)$ and $P_n(.,R/Wh)$ have for $n \geq n_0$ a positive lower bound
				  and thus the same sign at the boundary points of $E$.
				  
				  Ad b): Since by the last remark $P_n(x,R/\rho_\nu h)$ and $P_n(x,R/Wh)$ have the
				  same number of zeros in each gap the assertion has to be proved for weights of
				  the form $W=\rho_\nu$ only.

 		 If $ \delta_{j,n} = 1$ ,  
		 $j\in \{1,\ldots,l-1\}$, then it follows by \eqref{eq-26}, using the facts that 
		 $\abs{\phi(z,\infty)} > 1$ on $\bar\C \setminus E$ and that the modulus of each other
		 factor has a positive lower bound on $[a_{2j},a_{2j+1}]$, that
		 $|\psi _n| >1$ on $(a_{2j},a_{2j+1})$ if $n$ is sufficiently large. Thus by
		 \eqref{eq-34}  $P_n(. ,R/h\rho _\nu)$ has no zero in $(a_{2j},a_{2j+1})$.
		 
		 If $\delta _{j,n} = -1$ then $\psi _n $ has by \eqref{eq-26} exactly one 
		 zero in $(a_{2j},a_{2j+1})$, namely the simple zero $x_{j,n}$. Note that by \eqref{eq-230} and \eqref{eq-220}
		 $x_{j,n}$ can not coincide with a zero of $\rho _\nu$. Since $x_{j,n} \in 
		[a_{2j}+\tilde{\epsilon},a_{2j+1}-\tilde{\epsilon}]$, $\tilde{\epsilon}>0$, 
		it follows again by \eqref{eq-26} and $\abs{\phi(z,\infty)} > 1$ on $\bar\C
		\setminus E$ that $\psi _n$ is unbounded with respect to $n$ on every compact
		subset of $(a_{2j}, a_{2j} + \delta ] \cup [a_{2j+1} - \delta,
		a_{2j+1})$,$\delta > 0 $. Hence, using the facts that by \eqref{eq-34} $|\psi
		_n| = 1 $ at the boundary points of $E$ and that $\psi _n$ is real and
		continuous on $[a_{2j},a_{2j+1}]$ up to the poles $w_j$ we conclude that $\psi
		_n +1$ or $\psi _n -1$ has at least one simple zero in $(a_{2j},a_{2j+1})$.
		Since by Corollary \ref{cor1} either both $P_n$ and $Q_m$ have exactly one or no
		zero on $(a_{2j},a_{2j+1})$ it follows by \eqref{eq-34} that $P_n(.,R/h\rho _\nu)$ 
		has exactly one zero in $(a_{2j},a_{2j+1})$ which proves the
 		assertion.
%
		\end{proof}
		
		Thus the appearence of a zero of $P_n$ in a gap depends on the fact in which
		part of the Jacobian variety $\vec{\mathcal V}_n$ lies. Let us point out that
		the $\delta_{j,n}$'s from \eqref{eq-Nst-1} coincide with the $\delta_{j,n}$'s from
		Lemma\ref{lemma2}, relation\eqref{eq-2302}, if $W=\rho_\nu$.

 If the harmonic measures are rational then for Bernstein-Szeg\"o weights the
 number of zeros in the intervals can be determined easily.
 
 		\begin{prop}
			\label{prop1}
			 Assume that $\omega _j(\infty) = k_j/N$ for $j=1,\ldots,l-1$, where $k_j\in
			 \{1,\ldots,N-1\}$ and $N\in \N \setminus \{1\}$. Let $P_n$ be orthonormal to
			 $\mathbb P_{n-1}$ with respect to $d\mu_{R,\rho_\nu,\epsilon}$. Then for $\kappa =
			 0,\ldots,N-1$ and $m\in\N$ with $2(\kappa + mN) \geq \nu + l - \partial R -1$
			  \begin{equation}
			  	\#Z(P_{\kappa + (m+1)N}, E_j) = \#Z(P_{\kappa + mN}, E_j) + k_j, \quad 
			  	j=1,\ldots,l-1
				\label{eq-Nst170}
			  \end{equation}
			  \end{prop}
			  
			  \begin{proof}
				  Obviously for $m\in \N$, $\kappa \in \{0,\ldots,N\}$,
				  \begin{equation}
				  	(mN+\kappa)\vec{\omega}(\infty) = (mk_1+\frac{\kappa k_1}{N},\ldots, 
					mk_{l-1}+\frac{\kappa k_{l-1}}{N})
				  	\label{eq-Nst180}
				  \end{equation}
 				  which gives by \eqref{eq-Nst8} and Lemma \ref{lemma3.2} that the $x_{j,n}$'s and
 				  $\delta _{j,n}$'s from Lemma \ref{lemma3.1} and thus from Lemma \ref{lemma2}
 				  satisfy
				  \begin{equation}
				  	x_{j,\kappa + mN}= x_{j,\kappa + (m+1)N} \quad \text{and} \quad \delta_{j,\kappa
				  	+ mN}= \delta _{j,\kappa + (m+1)N}.
				  	\label{eq-**}
				  \end{equation}
				  which proves the
				  assertion. Let us note that \eqref{eq-**} has been shown already in
				  \cite{PehJAT,ref42} by different methods.
				  \end{proof}
				  Let us mention that the harmonic measures satisfy $\omega_j(\infty) = k_j/N$ for
				  $j=1,\ldots,l-1$ if and only if there exists a polynomial $\mathcal T$ of degree
				  $N$ such that $E=\mathcal T^{-1}_N([-1,1])$ (see \cite{Apt,Peh33,PehJAT}).

				  \begin{thm}
					  \label{HP}
					  Suppose that W satisfies the assumptions of Thm.\ref{Nst.Thm.1}.
					  Let $z_{j,n}(W) \in \mathcal R$, $j=1,\ldots,l-1$, be the points which solve the
					  Jacobi inversion problem, $k=1,\ldots,l-1$,
				 \begin{equation}
					 \begin{split}
		\frac{1}{2}\sum_{j=1}^{l-1}\int_{z_{j,n}^*(W)}^{z_{j,n}(W)} {\varphi _k}& = 
		\frac{1}{\pi i}\int_E \varphi _k^+ \log{|W|} \; - \; \frac{(2n + \partial R -
		l+1)}{2}\int_{\infty^-}^{\infty^+}{\varphi _k } \\ & \quad +
		\frac{1}{2}\sum_{j=1}^{l-1}\#Z(R,E_j)B_{kj} \quad \mod \text{periods}
		\end{split}
		\label{eq-Nst7F}
	          \end{equation}
	Furthermore let $z_k\in \mathcal R$, $k\in \{1, \ldots, l-1\}$, be such that
	$\pi (z_k)\in (a_{2k},a_{2k+1})$,\\ 
	where $\pi$ is the canonical projection of
	$\mathcal R$ on $\bar{\C}$. Then $\pi (z_k)$ is a limit point of zeros of
	$(P_{n_\kappa}(.,R/Wh))$ if and only if $z_k$ is a limit point of
	$(z_{k,n_\kappa}(W))$ with $z_k\in (a_{2k},a_{2k+1})^-$.
	\end{thm}
	
	\begin{proof}
		First let us prove the assertion for $W=\rho_\nu$. Put
		$z_{j,n}:=z_{j,n}(\rho_\nu)$ for $j=1,\ldots,l-1$ and let us observe first that by
		\eqref{eq-Nst7F}, Lemma \ref{lemma3.2}, \eqref{eq-neu} and \eqref{eq-40} the
		$x_{j,n}$'s, \ie the zeros of $g_{(n)}$, and the $\delta_{j,n}$'s from Lemma
		\ref{lemma3.1} and thus from Lemma \ref{lemma2}, are related to the $z_{j,n}$'s
		as follows:
		\begin{equation}
			\pi(z_{j,n}) = x_{j,n} \quad \text{and} \quad \delta _{j,n}= \pm 1\; \text{if}\;
			z_{j,n} \in (a_{2j},a_{2j+1})^\pm.
			\label{eq-HP1}
		\end{equation}

		Sufficiency. Since $z_{k,n_\kappa}\in (a_{2k},a_{2k+1})^-$ for $\kappa \geq \kappa_0$
		we have by \eqref{eq-HP1}
		\begin{equation}
			\delta_{k,n_\kappa}=-1 \quad \text{and} \quad \pi(z_{k,n_\kappa})=x_{k,n_\kappa}
			\label{eq-HP2}
			\end{equation}
			Relation \eqref{eq-26} implies that $\psi_{n_\kappa}(x_{k,n_\kappa})=0$, where
			$\psi_{n_\kappa}$ changes sign at $x_{k,n_\kappa}$. Since by \eqref{eq-26} and 
			$\abs{\phi(z,\infty)} > 1$ on $\bar\C \setminus E$
			$\psi_{n_\kappa}$ is unbounded with respect to $\kappa$ on compact subsets of
			$(a_{2k},a_{2k+1}) \setminus \{x_k\}$, where $x_k:=\pi(z_k)$, it follows that
			$\psi_{n_\kappa}$ takes on the value $-1$ on $(x_k-\varepsilon, x_k+\varepsilon)$,
			$\varepsilon>0$, if $\kappa$ is sufficiently large. By \eqref{eq-34} the
			sufficiency is proved.
 		   
 		   Necessity. Assume that $x_k:=\pi(z_k)$ is a limit point of zeros of the sequence
 		   $(P_{n_\kappa}(.,R/h\rho_\nu))$ and that $z_k$ is no limit point of
 		   $(z_{k,n_\kappa})$. Then as before $\psi_{n_\kappa}$ is unbounded with respect
 		   to $\kappa$ on $[x_k-\varepsilon, x_k+\varepsilon], \varepsilon >0$, which
 		   implies by \eqref{eq-24} that $P_{n_\kappa}(.,R/h\rho_\nu)$ is unbounded on this
 		   interval which is the desired contradiction.
		   
		   Next let us prove the assertion for the general case. Let $(\rho_\nu)$ be the
		   sequence given in \eqref{eq-Nst12} and \eqref{eq-Nst12Sch} and thus for $\nu \geq
		   \nu_0$,$n\geq 2\nu$
		   \begin{equation}
		   	z_{j,n}:= z_{j,n}(\rho_\nu) = z_{j,n}(W) \quad \text{for} \quad j=1,\ldots,l-1
		   	\label{eq-Hp3}
		   \end{equation}
		   Let $z_k\in(a_{2k},a_{2k+1})^-$ be a limit point of $(z_{k,n_\kappa})$ which
		   is, by what has been proved just, equivalent to the fact that $x_k:= \pi(z_k)$
		   is a limit point of zeros of $(P_{n_\kappa}(.,R/h\rho_\nu))$. Since, because of
		   $W\neq 0$ on $E$, the zeros of $\rho_\nu$ stay away from the boundary points
		   $a_{2k}$ and $a_{2k+1}$ for $\nu\geq \nu_0$ as well as does the zero
		   $x_{k,n_\kappa}=\pi(z_{k,n_\kappa})$ of $g_{(n_\kappa)}$, $ \kappa \geq
		   \kappa_0$, it follows by \eqref{eq-21neu} that

		   	\begin{equation}
		\abs{P_{n_\kappa}(z,R/h\rho_\nu)}\geq const > 0 \quad \text{for}\quad z\in
		\{a_{2k},a_{2k+1}\}
		\label{eq-*4}
	\end{equation}
	and by \eqref{eq-24} and \eqref{eq-26}
	\begin{equation}
		\abs{\frac{P_{n_\kappa}(z,R/\rho_\nu h)}{\phi (z,\infty)^{n_\kappa}}} \geq const
		>0 \quad \text{on}\; [a_{2k},a_{2k+1}]\setminus (x_k-\varepsilon,x_k+\varepsilon)
		\label{eq-HP4}
	\end{equation}
	  for $ \kappa \geq
	\kappa_0$, $\nu \geq \nu_0$. By \eqref{eq-Nst13} and \eqref{eq-*4} it follows that
		   	\begin{equation}
		\operatorname{sgn} P_{n_\kappa}(z,R/Wh)= \operatorname{sgn}
		P_{n_\kappa}(z,R/h\rho_\nu) \quad \text{for} \quad z\in \{a_{2k},a_{2k+1}\}
		\label{eq-*5}
	\end{equation}
	Hence, $P_{n_\kappa}(z,R/hW)$ has for each $\kappa \geq \kappa _0$ exactly
	one zero in $(a_{2k},a_{2k+1})$. By the asymptotic representation on $\C
	\setminus E$, see \cite{tomneu} or \cite{Nut-Sin},
				  \begin{equation}
		  	\frac{P_{n_\kappa}(z,R/Wh)}{\phi(z,\infty)^{n_\kappa}} = \frac{P_{n_\kappa}(z,R/\rho_\nu
		  	h)}{\phi (z,\infty)^{n_\kappa}} + O(\epsilon_{n_\kappa})\quad \text{with} \,\,
		  	\epsilon_{n_\kappa}\to 0 \,\, \text{as} \,\, \kappa \to \infty,
				   \label{eq-Nst160}
		  \end{equation}
		  and \eqref{eq-HP4} we conclude that every accumulation point of zeros of the sequence
		  $(P_{n_\kappa}(z,R/Wh))$ in $[a_{2k},a_{2k+1}]$ has to coincide with $x_k$ which
		  proves the sufficiency part.
		 
		 Concerning the necessity part let us assume that $x_k:=\pi(z_k)$ is a limit point
		 of zeros of $(P_{n_\kappa}(.,R/Wh))$ and that $z_k$ is no limit point of
		 $(z_{k,n_\kappa})$. Then by \eqref{eq-Hp3} and the same arguments as above
		 \begin{equation} \abs{\frac{P_{n_\kappa}(z,R/\rho_\nu h)}{\phi
		 (z,\infty)^{n_\kappa}}} \geq const >0 \quad \text{on}\; \;
		 [x_k-\varepsilon,x_k+\varepsilon], \varepsilon >0,
		 \label{eq-*40}
		 \end{equation}
		 which gives by \eqref{eq-Nst160} a
		 contradiction.
		  \end{proof}

          \begin{thm}
			  \label{DThm}
			  Let W satisfy the assumptions of Thm.\ref{Nst.Thm.1}.
			  \begin{enumerate}
				  \item [a)] Suppose that $1, \omega _1(\infty),\ldots, \omega_{l-1}(\infty)$ are
				  linearly independent over the rationals. Let $x_{j_\mu}\in (a_{{2j}_\mu},
				  a_{{2j}_\mu+1})$, $\mu = 1,\ldots, l' $, $l'\leq l-1$, be given. Then there
				  exists a sequence $(n_\kappa)$ such that each $x_{j_\mu}$ is a limit point of
				  zeros of $(P_{n_\kappa}(z,R/Wh))$ and $P_{n_\kappa}(z,R/Wh)$, $\kappa \in \N$,
				  has no zero in the other gaps $[a_{2j},a_{2j+1}], \, j\in \{1,\ldots,l-1\}
				  \setminus \{j_\mu :\mu = 1,\ldots,l'\}$. \item[b)] Suppose that $\omega
				  _j(\infty) = k_j/N$ for $j=1,\ldots,l-1$, where $k_j\in \{1,\ldots,N-1\}$ and
				  $N\in \N \setminus \{1\}$. Then $(P_n(z,R/Wh))$ has at most $N+2$ accumulation
				  points in each gap $[a_{2j},a_{2j+1}]$, $j=1,\ldots,l-1$.
			  \end{enumerate}
			  \end{thm}
 
	\begin{proof}
		Ad a) Put $\vec{x}=(x_1,\ldots,x_{l-1})$, where $x_j\in(a_{2j},a_{2j+1})$ is arbitrary for
		$j\in\{1,\ldots,l-1\}\setminus \{j_1,\ldots,j_{l'}\}$ 
	 and put
			\begin{equation}
		 \delta_j=\begin{cases}
		-1& \text{for $j\in \{j_1,\ldots,j_{l'}\}$,}\\
		+1&\text{for $j\in\{1,\ldots,l-1\}\setminus \{j_1,\ldots,j_{l'}\}$.}
		\end{cases}
		\label{eq-Nst100}
		\end{equation}
		Since the harmonic measures are rationally independent it follows from Kronecker's
		Lemma, see e.g.\cite[p.23]{Hla}, and \eqref{eq-Nst8} that there exists a sequence
		$(n_\kappa)$ of natural numbers such that for $ k = 1,\ldots, l-1$,
		\begin{equation}
			\begin{split}
 - &\frac{(2n_\kappa + \partial R - l+1)}{2} \int_{\infty^-}^{\infty^+}\varphi _k +
				  \frac{1}{\pi i} \int_E \varphi _k \log {|W|} \\
				  &+\sum_{j=1}^{l-1}(\frac{\# Z(R,E_j)}{2}+N_{j,n_\kappa})B_{kj} \quad \to \quad
				  \frac{1}{2} \sum_{j=1}^{l-1} \delta _j \int_{x_j^-}^{x_j^+}\varphi _k =
				  \frac{1}{2} \sum_{j=1}^{l-1} \int_{z_j^*}^{z_j}\varphi _k
		   \end{split}
		   \label{eq-Nst13E}
	   \end{equation}
as $\kappa \to \infty$, where $N_{j,n_\kappa}\in\Z$, $z_j:=x_j^{\delta_j}$. Now let
$z_{j,n_\kappa}$ be the solutions of the Jacobi inversion problem
\eqref{eq-Nst7F} for $n_\kappa$, $\kappa \in \N$. Then it follows by Lemma
\ref{lemma3.2} and continuity arguments that
	  				  \begin{equation}
				  z_{j,n_\kappa} \to z_j \quad \text{as}\quad  \kappa \to \infty.
				  	\label{eq-*}
				  \end{equation}
				  Since $z_j\in \mathcal R^{\delta_j}$, $j=1,\ldots,l-1$, part a) follows by Theorem 
				  \ref{HP}.
				  
				  Ad b) In view of \eqref{eq-Nst180} and Lemma \ref{lemma3.2} the solutions of
				  \eqref{eq-Nst7F} satisfy
				  \begin{equation}
				  	z_{j,\kappa + mN}(W)= z_{j,\kappa + (m+1)N}(W)
				  	\label{eq-D1}
				  \end{equation}
				  for $j=1,\ldots,l-1$, $n,m\in \N$ and $\kappa = 0,\ldots, N-1$ which gives by
				  Theorem \ref{HP} the assertion.
				  \end{proof}
				  
				  If we put $l'=0$ in Theorem \ref{DThm}a) we obtain a recent result of S.P. Suetin
				  \cite{Sue} who has shown (even for arcs loosley speaking) that there exist a
				  subsequence $(n_\kappa)$ such that the zeros of $(p_{n_\kappa})$ accumulate on
				  $E$ only.
	  
		With the help of similar ideas used in this paper we are also able to give
		corresponding results for polynomials orthogonal with respect to varying
		exponential weights which are of particular interest in connection with random
		matrices (see \cite{Deietal2,Dei1,Dei}) as well as for polynomials orthogonal or
		extremal on more general sets in the complex plane. The results will be given
		elsewhere.

\end{document}